# Solute effects on interfacial dislocation emission in nanomaterials: nucleation site competition and neutralization

*Valery Borovikov[1], Mikhail I. Mendelev[1] and Alexander H. King[1,2]*

[1]Division of Materials Sciences and Engineering, Ames Laboratory, Ames, IA 50011

[2]Department of Materials Science and Engineering, Iowa State University, Ames, IA 50011

Interfacial nucleation is the dominant process of dislocation generation during the plastic deformation of nano-crystalline materials. Solute additions intended to stabilize nano-crystalline metals against grain growth, may segregate to the grain boundaries and triple junctions where they can affect the process of the dislocation emission. In this Letter we demonstrate that the effect of solute addition in a nano-crystalline material containing competing solute segregation sites and dislocation sources can be very complex due to different rates of segregation at different interfaces. Moreover, at large concentrations, when the solutes form clusters near the grain boundaries or triple junctions, the interfaces between these clusters and the matrix can introduce new dislocation emission sources, which can be activated under lower applied stress. Thus, the strength maximum can occur at a certain solute concentration: adding solutes beyond this optimal solute concentration can reduce the strength of the material.

Grain boundaries (GB) serve as obstacles to dislocation motion in conventional materials[1] and as their density increases yield strength increases along with decreasing grain size, according to the classical Hall-Petch relationship. A maximum strength is achieved for very small grain sizes, however, and further reductions in grain size can result in softening due to the activation of interfacially-mediated plasticity mechanisms such as grain boundary sliding and grain rotation [2, 3]. Nano-crystalline materials, also suffer from a strong tendency for grain growth and coarsening even at relatively low temperatures, as a result of the extreme driving forces provided by the interfacial energy. These tendencies can be decreased by small solute additions [4], which tend to segregate at the grain boundaries, reducing their mobility and suppressing grain boundary mediated plasticity mechanisms[5-9]. However, solutes may also affect the process of dislocation emission from GBs[10] and triple junctions (TJ), which is one of the key mechanisms of plastic deformation at nanoscale grain sizes [11-13]. As grain size, stress and temperature vary, several mechanisms of plastic deformation can be active simultaneously, competitively and/or co-operatively, making it difficult to isolate and study any individual deformation mechanism. This explains why the effect of solute atoms on dislocation nucleation from GBs has not been systematically studied. We have recently used atomistic simulations to provide direct evidence of the strong effect of solutes segregated at a GB on dislocation nucleation and yield stress under tensile loading [14], using a simple bi-crystal geometry, for which the nucleation and propagation of dislocations from a GB is the only possible mechanism of plastic deformation. While the bi-crystal geometry provides a very convenient way to study the dislocation nucleation from a particular GB, some potentially important aspects of the deformation behavior of real materials are not taken into account when such a geometry is employed[15]. In particular, in a nano-crystalline material under applied stress, dislocation slip can be initiated from multiple competing sources located at GBs and TJs. Therefore, the dependence of the yield stress on the solute concentration can be more complex than revealed by the results we obtained using the bi-crystal geometry in [14].

In the present study we have employed a more complex, yet still tractable multi-grain geometry, previously used in our study of the dislocation nucleation in pure nano-twinned materials [16]. The simulation cell contains different grain and twin boundaries and their triple junctions (see Fig. S1 of the Supporting Information). This geometry resembles the microstructure of sputtered thin films produced and tested experimentally [17, 18]. The simulation cell contains one type of asymmetric tilt grain boundaries (ATGB), two distinct types of symmetric tilt grain boundaries (STGB), coherent twin boundaries (CTB) and four distinct types of TJs between the GBs and CTBs. Dislocation emission from grain boundaries and triple junctions is the only mechanism of plastic deformation in this system for tensile loads applied in the direction normal to the grain boundaries. Thus the chosen geometry allows us to study how different dislocation nucleation sources compete with each other. In the present work, we studied Ag with varying Cu additions. The atomic interactions were described by an embedded-atom method (EAM) potential developed in [19]. Due to the low solubility of Cu in Ag [20], the solutes have a strong tendency for segregation at the GBs and TJs.

All simulations were carried out using the LAMMPS simulation package [21] and the visualization of the simulation snapshots was performed using the software package OVITO [22]. The preparation of the pure Ag model was described in details in [16] and the way to introduce the Cu solute atoms and equilibrate the simulation cell using a hybrid Monte Carlo/molecular dynamics (MC/MD)[23] simulations was described in [14].

Prior to tensile loading, all simulation cells were equilibrated at T=300 K and zero applied stress. The tensile loading simulations were carried out with a constant engineering strain rate of $10^8$ s$^{-1}$. The deformation was applied in the x-direction (normal to the average GB orientation), while the stresses in the other two directions were kept zero. Stress-strain curves obtained for different solute concentrations are shown in Fig. 1. In all cases, the stress first increases with increased applied strain until the emission of the first dislocation. The value of the stress at that moment was considered as the yield stress in the present study. The dislocations are emitted from the multiple different competing sources, located at the GBs and TJs. Figure 2 shows the examples of dislocation emission from a GB and a TJ.

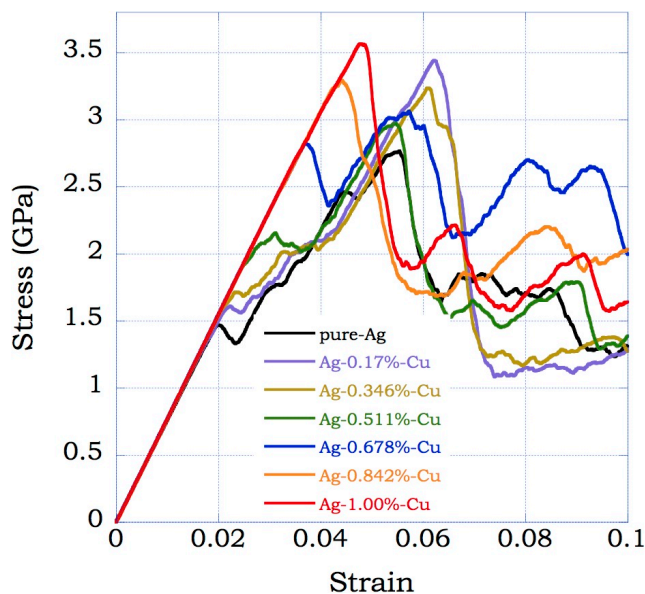

**Figure 1.** Stress-strain curves obtained using different solute concentrations (at.%).

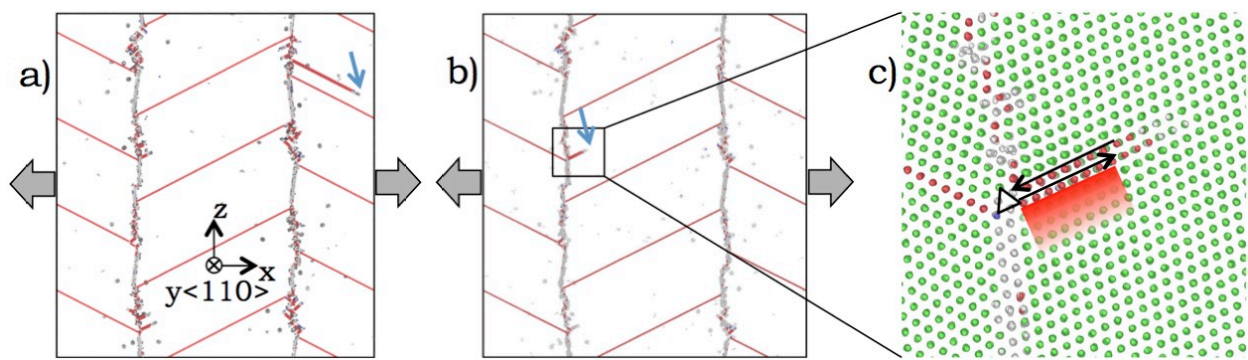

**Figure 2.** Examples of dislocation emission from competing sources. The large grey arrows indicate the direction of the applied deformation. The atoms are colored according to the Common Neighbor Analysis (CNA)[24, 25]. The color-coding is as follows: green – FCC, red – HCP, grey – other. The small blue arrows highlight the dislocations emitted from a) GB, and b) TJ (fcc (bulk) atoms are not shown). c) The enlargement of the rectangular segment (all atoms are shown) from b). The triangular structural unit, which serves as a dislocation emission source,

is highlighted. The arrows schematically indicate the direction of motion for two atomic planes in the course of dislocation emission.

Figure 3a shows that the dependence of the yield stress on the solute concentration is not monotonic. There are three well-defined regimes. At low solute concentrations, the yield stress slowly increases with increasing solute concentration. When a critical solute concentration is reached (~0.5%), we observe a more rapid increase of the yield stress with increasing concentration of the solutes, and the maximum yield strength is reached at the optimal solute concentration (~1%). At concentrations above this level, we observe a decrease in the yield stress.

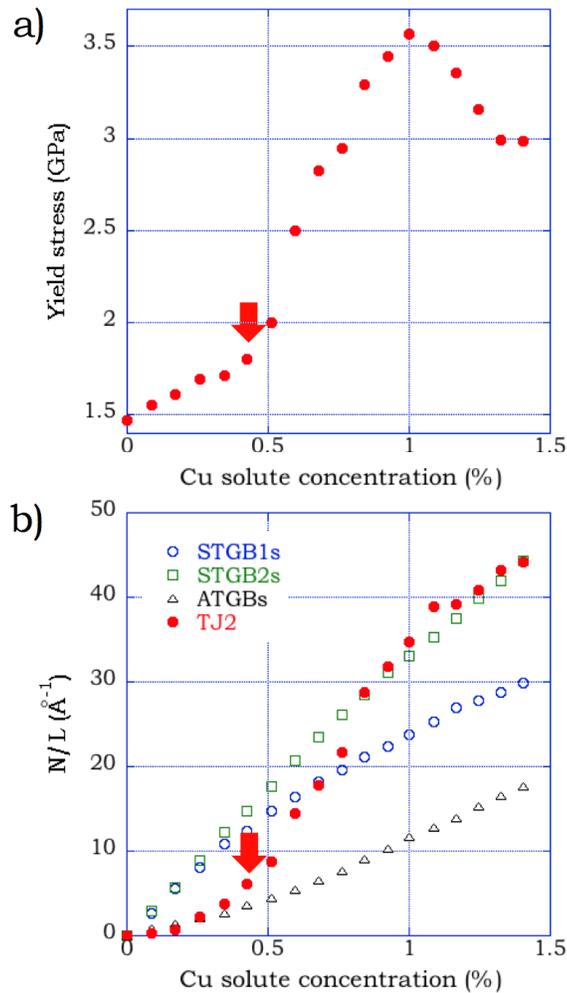

**Figure 3.** a) The dependence of the yield stress on solute concentration (at.%). b) The number of solutes per unit of length at different GBs and a TJ. The red arrows indicate the solute concentration at which the yield stress begins to increase rapidly. This is associated with the accelerated segregation of the solutes at the triple junctions, and the corresponding neutralization of the last active dislocation sources located there.

In order to explain the dependence of the yield stress on solute concentration we investigated the solute concentrations at GBs and TJs, as follows. The regions near the GBs and TJs were divided in slices perpendicular to the average GB orientation (the yz plane). The example is shown in Fig. S2 of the Supporting Information. The number of solute atoms in each slice was determined and normalized by the slice thickness (in the z-direction). The results obtained for GBs and TJ2 (intersection of STGB1, ATGB and CTB) are shown in Fig. 3b. This figure vividly shows that the segregation differs for the individual GBs and the TJ, consistent with experimental observations of TJ segregation [26]. Significantly, the rates of segregation also vary. At lower solute concentrations (<0.5%) the solutes segregate preferentially at the symmetric tilt grain boundaries (STGB1 and STGB2). This leads to neutralization of the active dislocation sources located at them, paralleling the effect we observed in the bi-crystal geometry simulations [14]. The asymmetric GBs and the triple junctions receive much less solutes, compared to the symmetric tilt grain boundaries (Fig. 3b). As a result, the active sources located at the TJs continue to generate dislocations under relatively low applied stresses, minimizing the impact of solute additions to the symmetric boundaries. However, as soon as all of the most energetically favorable for segregation sites at the symmetric tilt grain boundaries are filled, solute atoms begin to segregate at the triple junctions. This shuts down the active dislocation sources located at the TJs and the yield stress rapidly increases with further increased solute concentration (see Fig. 3).

One might expect that once all the active dislocation sources are neutralized by the solute atoms, the yield stress would reach a constant value and further increase of the solute concentration would not affect it anymore. Figure 3a shows that it is not the case: the yield stress reaches a maximum at ~1% of Cu concentration and then drops at higher solute concentrations. To explain this phenomenon we will take a closer look at the processes happening at the triple junction (TJ2) where emission of the first dislocation was observed for various solute concentrations. Fig. 2c, shows a region of compressive stress (highlighted in red) just below the source, where the undersized Cu atoms tend to segregate. Figure 4 shows in detail the evolution of stress fields at the triple junction (shown in Fig. 2c) with increasing solute concentration. Each pair of images in Fig. 4 shows the composition (left) and the atomic stresses (right) after energy minimization at zero applied stress. The blue arrows indicate the directions in which the atomic planes slide in the course of the dislocation emission. The material below the atomic plane indicated by the blue arrow, pointing to the right, shifts away from the compressive region and the material above the atomic plane indicated by the blue arrow, pointing to the left, shifts toward the region under hydrostatic tension. There is a very sharp boundary between the compressive and tensile regions in the pure Ag. The undersized Cu solutes segregate in the compressive region just below the initial source and make the boundary between the compressive and tensile regions much less pronounced, leading to reduced dislocation emission potential at the initial source. As shown in Fig. 4c, the boundary between the compressive and tensile regions eventually disappears and the initial source shuts down. However, as solute atom segregation extends below the TJ a new boundary between compressive and tensile regions emerges below the initial source. As a result, a different dislocation source activates at higher solute concentrations, two planes below the one that operates at lower concentrations. The red arrows in Figs. 4d, 4e schematically indicate the directions in which the atomic layers slide as a result of dislocation emission from the new source. The yield stress reaches its peak value at the intermediate stage when the initial source is completely neutralized and the new source has not been fully activated. Subsequently, as the solute cluster grows in size with increasing solute concentration, it becomes easier and easier to emit a dislocation from the new source, which explains the observed decrease in the yield stress at higher solute concentrations (>1%). We have observed the activation of several new dislocation sources (similar to the one discussed above) at higher solute concentrations in our simulations. In all cases the new sources are formed at the interface between the solute cluster and the matrix atoms.

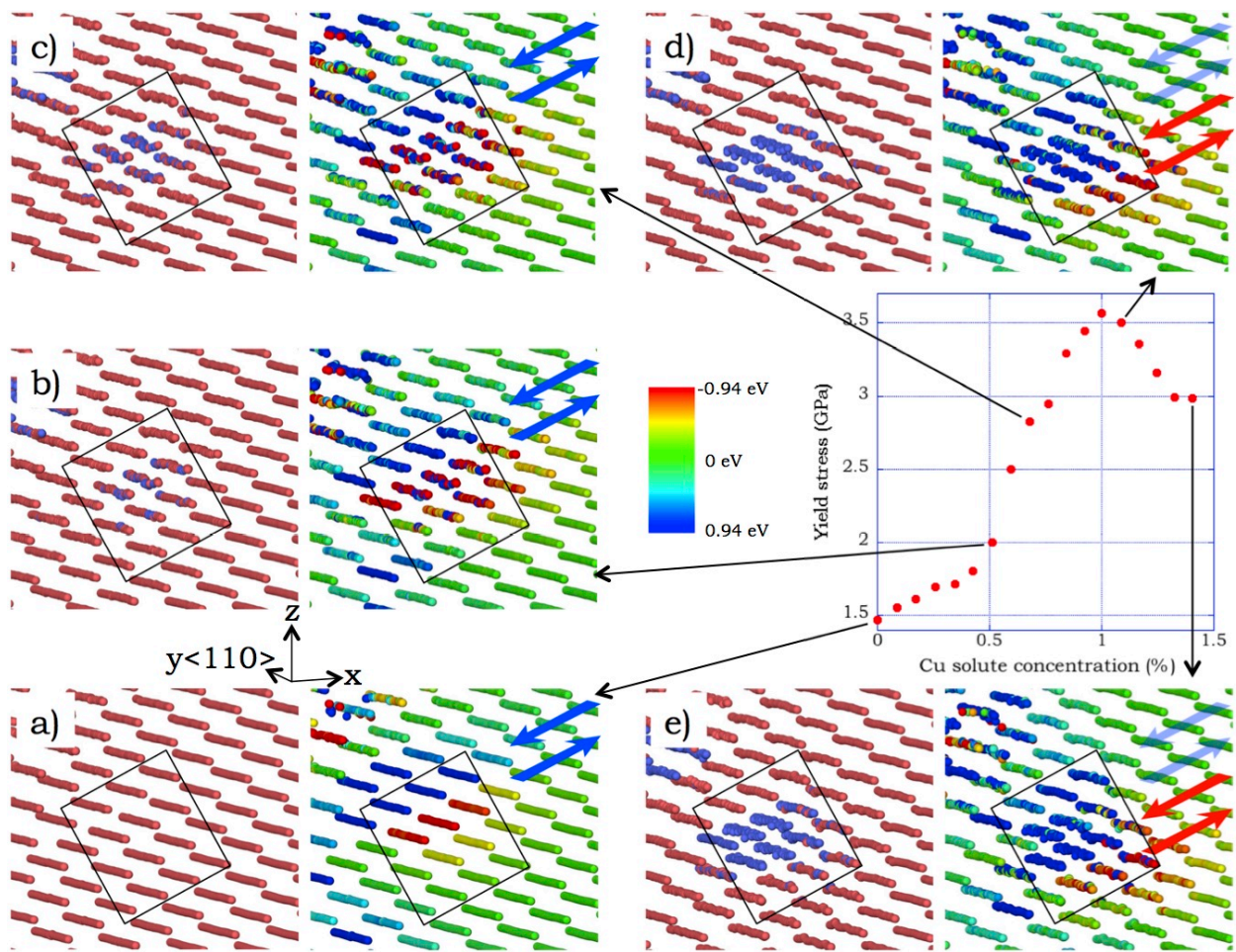

**Figure 4.** The evolution of the atomic stresses with solute concentration (at.%) in the region near the triple junction (TJ2). Each pair of images is colored according to the composition (left) and the trace of the atomic stresses (right). The blue arrows schematically indicate the directions in which the atomic planes slide in the course of the dislocation emission from the initial source at the triple junction. The red arrows schematically indicate the directions in which the atomic planes slide in the course of the dislocation emission from the new source (at high Cu concentrations). The rectangular box is located at the same position on each image, helping to track the changes in the stress fields near the triple junction with increasing solute concentration.

Our simulations indicate that solute atoms can have a very strong effect on dislocation nucleation and, therefore, on the yield stress in nanoscale materials, where interfacial nucleation is the dominant process of dislocation generation. The effect is non-linear and can be weak at low solute concentrations because some of the sources (more energetically favorable for segregation of the solutes) are neutralized first, while the other sources (less favorable for the segregation) are still able to emit dislocations under relatively low applied stresses. At higher solute concentrations, the addition of a small amount of solutes can considerably increase the yield stress because the solute atoms begin to segregate to the remaining active dislocation sources. However, further increases of the solute concentration can lead to decreasing yield stress due to the formation of solute clusters at the grain boundaries and triple junctions. At higher solute concentrations, the interfaces between these clusters and the matrix introduce new dislocation emission sources, which can be activated under lower applied stresses. Thus, when the solute atoms are added in order to make material stronger, a maximum strength may be achieved at some optimal solute concentration. Adding solute in excess of this optimal concentration may lead to a decrease of the yield strength.

**Acknowledgement**

This work was supported by the U.S. Department of Energy, Office of Science, Basic Energy Sciences, Materials Science and Engineering Division. The research was performed at Ames Laboratory, which is operated for the U.S. DOE by Iowa State University under contract # DE-AC02-07CH11358.

# Supporting Information

# Solute effects on interfacial dislocation emission in nanomaterials: nucleation site competition and neutralization

*Valery Borovikov[1], Mikhail I. Mendelev[1] and Alexander H. King[1,2]*

[1]Division of Materials Sciences and Engineering, Ames Laboratory, Ames, IA 50011

[2]Department of Materials Science and Engineering, Iowa State University, Ames, IA 50011

The simulation cell used in the present study (see Fig. S1) contains one type of asymmetric tilt grain boundary (ATGB), two distinct types of symmetric tilt grain boundaries (STGB), coherent twin boundaries (CTB) and corresponding triple junctions (TJ). The concentration of solutes was calculated for STGB1, STGB2, ATGB and four TJs (TJ1, TJ2, TJ3, TJ4). In order to do this the regions near the GBs and TJs were divided in slices perpendicular to the average GB orientation (the yz plane). The example is shown in Fig. S2. The number of solute atoms in each slice was determined and normalized by the slice thickness (in the z-direction). Since the system contains multiple GBs of the same type, the solute concentration (see Fig. 3b of the main text) shown for the GBs is the average concentration over the particular type of the GB. The concentration of solutes at four TJs indicated in Fig. 1S is shown in Fig. S3, while the Fig. 3b in the main text shows the result for a single TJ (TJ2). We note that the system contains other triple junctions, which do not serve as easily activated dislocation sources.

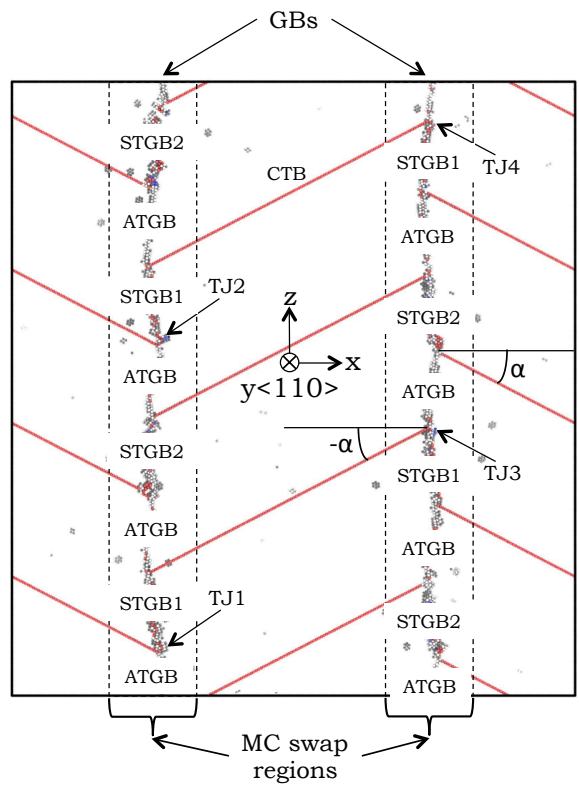

**Figure S1.** Simulation cell with atoms colored according to the Common Neighbor Analysis (CNA)[1, 2] and fcc (bulk) atoms are not shown. The color-coding is as follows: red – HCP, blue – BCC, grey – other. During deformation a strain was applied in the x-direction while the stresses in the y and z – directions were kept zero by applying a barostat.

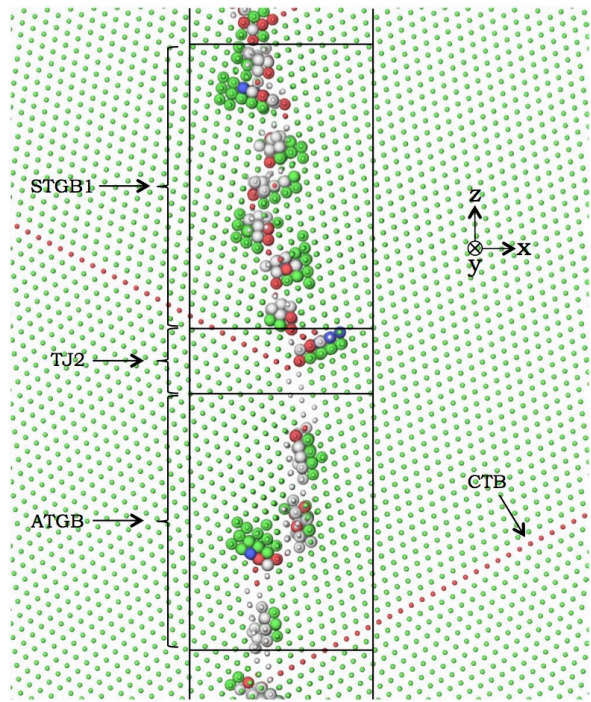

**Figure S2.** The examples of the regions near the GBs and TJs used in the calculation of solute concentration are shown schematically. The atoms are colored according to the Common Neighbor Analysis (CNA)[1, 2]. The corresponding total Cu solute concentration in the system is ~0.68%. The color-coding is as follows: green – FCC, red – HCP, blue – BCC, grey – other.

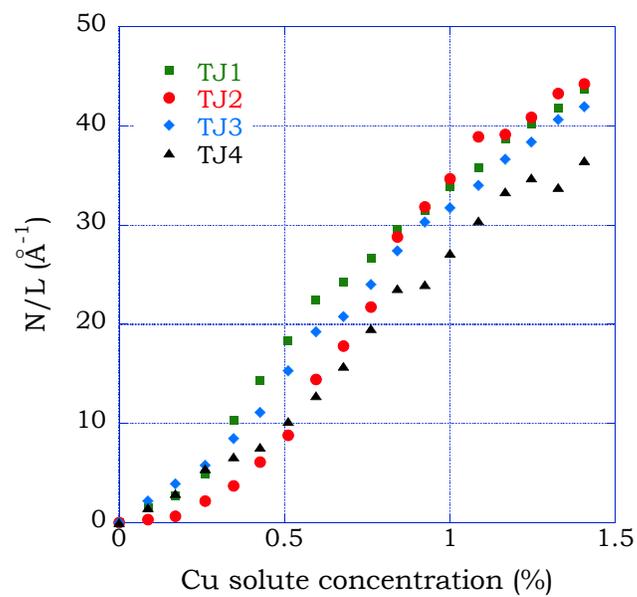

**Figure S3.** The number of solutes per unit of length for the triple junctions.